\documentstyle[pra,aps,multicol,epsfig]{revtex}

\input epsf.tex

\newcommand{\be}{\begin{equation}}
\newcommand{\ee}{\end{equation}}
\newcommand{\ba}{\begin{eqnarray}}
\newcommand{\ea}{\end{eqnarray}}
\newcommand{\ban}{\begin{eqnarray*}}
\newcommand{\ean}{\end{eqnarray*}}

\newcommand{\sandwich}[3]{\mbox{$ \langle #1 | #2 | #3 \rangle $}}
\newcommand{\ket}[1]{\mbox{$ | #1 \rangle $}}
\newcommand{\bra}[1]{\mbox{$ \langle #1 | $}}

\newcommand{\compl}{\begin{picture}(8,8)\put(0,0){C}\put(3,0.3){\line(0,1){7}}\end{picture}}

\newcommand{\one}{\leavevmode\hbox{\small1\normalsize\kern-.33em1}}

\begin{document}

\title{Violation of Bell's inequalities implies distillability for $N$ Qubits}
\author{Antonio Ac\'\i n$^1$, Valerio Scarani$^1$ and Michael M. Wolf$^2$}
\address{
$^1$Group of Applied Physics, University of Geneva, 20, rue de
l'Ecole-de-M\'edecine, CH-1211 Geneva 4, Switzerland\\
$^2$Institut f\"{u}r Mathematische Physik, TU Braunschweig,
Mendelssohnstrasse 3, 38106 Braunschweig, Germany}
\date{\today}
\maketitle

\begin{abstract}
We consider quantum systems composed of $N$ qubits, and the family
of all Bell's correlation inequalities for two two-valued
measurements per site. We show that if a $N$-qubit state $\rho$
violates any of these inequalities, then it is at least bipartite
distillable. Indeed there exists a link between the amount of
Bell's inequality violation and the degree of distillability.
Thus, we strengthen the interpretation of Bell's inequalities as
detectors of useful entanglement.
\end{abstract}

\begin{multicols}{2}

\section{Introduction}

Quantum mechanics predicts remarkable correlations between the
outcomes of measurements on sub-systems (particles) of a composed
system. This prediction is a consequence of the interplay between
the superposition principle and the tensor product structure of
quantum mechanics, that allows to build states that cannot be
written as probabilistic mixtures of products of states of each
sub-system. Such states are called non-separable or {\em
entangled}. This paper concerns the characterization of
entanglement for multi-partite systems. This is a complex problem.
The variety of known partial results suggest that there may not be
a unique characterization of entanglement. We show a close
connection between two features of entangled states, {\em
distillability} and {\em Bell's inequality violation} in systems
of $N$ quantum bits (qubits). But first, let's begin by reviewing
the notions of separability, distillability and quantum
non-locality.

{\em Separability.} A pure state $\ket{\Psi}$, shared by $N$
partners is separable when it can be written as the tensor product
of pure states of each subsystem, i.e.
$\ket\Psi=\ket{\psi_1}\otimes\ldots\otimes\ket{\psi_N}$. Thus, for
these states there are no correlations between the different
sub-systems. For the mixed-state case, a density matrix, $\rho$,
is separable when it can be expressed as the convex combination of
separable pure states, $\rho=\sum_i p_i
\ket{\Psi^s_i}\bra{\Psi^s_i}$, where $\ket{\Psi^s_i}$ are product
states $\forall i$. States that are not separable are called
entangled. It is clear from these definitions that if the initial
state of the system is separable and the partners are allowed to
use only local operations and classical communication (LOCC), then
definitely no entangled state can be prepared.

{\em Distillability} \cite{dist} An entangled state $\rho$ is
distillable if some partners sharing arbitrary many copies of the
state can produce a few maximally entangled states \cite{maxent}
using only LOCC. Once two partners share a maximally entangled
state, they can run quantum communication protocols, like
teleportation \cite{telep} or quantum key distribution \cite{qkd}.
Thus distillability is a measure of the ``usefulness" of a state:
if a state is distillable, we can extract from many copies of it
that kind  of entanglement needed to implement quantum
communication protocols.

A necessary (and conjectured not sufficient \cite{DCLB,NPTBES})
condition for distillability in the bipartite case is the
non-positivity of the partial transpose (NPPT): the partial
transpose of $\rho$ \cite{PT} must have at least one negative
eigenvalue \cite{bound}.

{\em Quantum non-locality.} If we focus on their preparation, all
entangled states have some form of quantum non-locality, in the
sense that they cannot be prepared without using either non-local
quantum operations or quantum channels. However in this article we
will use the term quantum non-locality for denoting the
impossibility of reproducing the correlations observed in some
quantum states of a multi-partite system by means of local
variable theories (LV).

Assume that a quantum state $\rho$ is prepared at a given location
and distributed among $N$ partners at different locations. The
partners can apply a LOCC protocol that transforms $M$ copies of
it into a new state, $\rho'$, with some probability. Then they
perform a sequence of measurements on it. We will call a state
{\it local} if the statistics of the outcomes can be reproduced by
a LV model for all these measurement and LOCC protocols. If there
exists a protocol such that the outcomes cannot be reproduced by
any LV model, we say that the state exhibits some {\em quantum
non-locality}. The last step for the detection of quantum
non-locality consists on checking if the measurement outcomes
violate some constraints, known as Bell's inequalities
\cite{Bell,review}, that any local variable model satisfies. If
the state $\rho'$ violates a Bell's inequality it is clearly
quantum non-local. We will also say $\rho$ to be quantum non-local
since it can be transformed into $\rho'$ without non-local quantum
resources.


No necessary and sufficient conditions for quantum non-locality
are known; in particular, nobody knows whether all entangled
states are non-local. Since the work of Werner \cite{werner} (and
its extension by Barrett \cite{Barrett}), it has been known that
there exist entangled states that do not violate any Bell's
inequality. This does not mean that they cannot violate a Bell's
inequality after LOCC filtering protocols \cite{popescu}. However,
it is conjectured that some of the states discussed in
\cite{werner,Barrett} are not distillable. Hence, they may not
exhibit some quantum non-locality although they are entangled.

We have completed our review of distillability and quantum
non-locality. Let's consider a first link between these two
notions. From the previous definitions it follows that if $\rho$
is distillable then it is also quantum non-local. In fact, if
$\rho$ is distillable, then the distilled maximally entangled
states violate a Bell's inequality for suitable measurements and
in this way reveal the quantum non-locality of the initial state.

The simplest composite scenario corresponds to a two-qubit system.
There, the Bell's inequality due to Clauser, Horne, Shimony and
Holt \cite{CHSH} plays an essential role. In the case of two
two-outcome measurements per site, it provides a necessary and
sufficient condition for the existence of a local variable model
\cite{Fine}. Moreover, all entangled states of two qubits can be
distilled to the singlet form \cite{horo}. In summary, the physics
of entanglement for two qubits conveys the intuitive idea that all
entangled states shall ultimately show some form of quantum
non-locality and can be used as resources in quantum information
protocols. However, as soon as one moves to more complex quantum
systems, these results are not yet proven or are proven to be
wrong! The most surprising feature is the existence of {\em bound
entangled states}, that are states that are entangled but cannot
be distilled \cite{bound}. Such states \cite{note0} seem to be
``useless" for quantum information protocols: the entanglement
that has been used to prepare them \cite{irrev} cannot be
recovered (at least, without the help of additional quantum
resources \cite{activation}).

\begin{center}
\begin{figure}
\epsfxsize=8cm \epsfbox{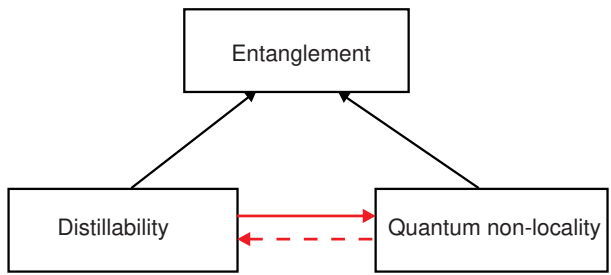} \vspace{.3 cm}\caption{The
figure summarize the known implications between entanglement,
distillability and quantum non-locality. If a state is distillable
or it cannot be described by a LV model, then it is entangled.
Bound entangled states show that not all entangled states are
distillable. Whether all entangled states have quantum
non-locality is an open question. According to our definitions,
distillability implies quantum non-locality. In this article we
explore whether the converse is also true (dashed line) in systems
of $N$ qubits.} \label{connect}
\end{figure}
\end{center}

In this paper, we study the link between Bell's inequalities and
distillability on a quantum system composed of $N$ qubits.
Recently, a large family of Bell's inequalities for multipartite
systems has been completely characterized; thus we can study the
link with distillability on a by now well-known mathematical
structure. We will in particular discuss the situation where the
$N$ qubits are shared by less than $N$ distant parties (Section
\ref{secqubits}). In Section \ref{secres} we then present the main
results, that can be summarized as follows: (i) If a state of $N$
qubits violates any of the inequalities in the family that we
consider, then one can always find at least two subgroups that can
distill a singlet if quantum communication is allowed within each
group; (ii) The amount of the violation of the inequality provides
an upper bound to the size of the subgroups; the higher the
violation, the smaller the subgroups; (iii) In particular, if the
violation of the inequality is sufficiently large, {\em each pair}
of qubits can distill a singlet, i.e. there is no need to
establish a quantum channel. Using teleportation, this implies
that any quantum state of $N$ qubits can be distilled (``full
distillability") by means of operations which are local with
respect to all $N$ tensor factors.

\section{$N$ qubits as a multi-composed system}
\label{secqubits}

\subsection{Multi-qubit entanglement and distillability}

The classification of entanglement for multi-composed systems
(here we stick to the case of $N$ qubits) is not an easy task.
Indeed, it is not even know what the fundamental types of
pure-state entanglement are \cite{AVC}. In a qualitative way, some
extreme cases are easily found: on one side, full product states,
$\ket{\psi_1} \otimes...\otimes \ket{\psi_N}$; on the other side,
fully entangled states, i.e. pure states that are bipartite
entangled with respect to any splitting of the parties into two
groups. A known example of the latter kind of states is the
$N$-qubit GHZ state $\ket{\mbox{GHZ}}_N=
1/\sqrt{2}\big(\ket{0}^{\otimes N}+\ket{1}^{\otimes N}\big)$. But
in-between these extreme cases the zoology is extremely rich, and
has been the object of several studies. One can tentatively define
a {\em degree of entanglement}. A state like
$\ket{\mbox{GHZ}}_n\otimes \ket{0}^{N-n}$ is clearly $n$-partite
entangled. Extending this definition to density matrices, one can
say that $\rho$ has $n$-partite entanglement if, in any possible
decomposition of $\rho$ into mixtures of pure states, there is at
least one state that is $n$-partite entangled ($n$-partite
entanglement was needed in the state preparation).

Now, what does it mean for a multi-partite state to be
distillable? As we wrote in the Introduction, in the bipartite
case, a state $\rho$ is distillable when out of possibly many
copies of it the two parties can extract a two-qubit maximally
entangled state by LOCC. It is not completely evident how to
extend this definition to a multi-partite scenario. We proceed in
two steps.

Let us first define {\em full distillability}: a quantum state
shared by $N$ parties is $N$-partite (``fully") distillable when
it is possible to distill singlet states between any pairs of
parties using LOCC. This is equivalent to demand that any truly
$N$-partite entangled state, in particular an $N$-qubit GHZ state,
can be obtained by LOCC. Indeed, once all the parties are
connected by singlets, one of them can prepare locally any of
these states and send it to the rest by teleportation.  On the
other hand if the parties share an $N$-partite pure entangled
state, there exist local projections such that $N-2$ qubits
project the remaining two parties into a bipartite entangled pure
state \cite{PR}, which is always distillable.

Consider now a state $\rho$ of $N$ qubits which is {\em not}
$N$-partite distillable. Still it may happen that, if some of the
qubits join into several groups (or establish quantum channels
between them), then the state, now shared by $L<N$ parties, is
$L$-partite distillable \cite{DC}. That is: without using any
global quantum operation between them, but just using LOCC, each
two among the $L$ parties can distill a singlet.

Thus, it is possible to estimate the {\em degree of
distillability} in an $N$-qubit state by means of the minimal size
of the groups the parties have to create in order to distill
singlets, i.e. pure-state entanglement, between them. For
instance, in the situation shown in figure \ref{figthm2} (see
below) this minimal size is equal to three. We have two extreme
cases in this classification: (i) the parties can perform all the
operation without joining, and then the state is fully distillable
as was defined above, or (ii) they have to join into two groups,
and then the state is said to be bipartite distillable \cite{DC}.
All the other cases lie between these two possibilities.

\subsection{The family of Bell's inequalities}

A complete set of Bell correlation inequalities for $N$-partite
systems  was found by Werner and Wolf \cite{WW}, and independently
by Zukowski and Brukner, in \cite{ZB}. Every local observer,
$i=1,\ldots,N$, can measure two dichotomic observables, $O^1_i$
and $O^2_i$, whose outcomes are labelled $\pm 1$. Thus, after many
rounds of measurements, all the parties collect a list of
experimental numbers, and they can construct the corresponding
list of correlated expectation values,
$E(j_1,j_2,\ldots,j_N)=\langle O^{j_1}_1\otimes
O^{j_2}_2\otimes\ldots\otimes O^{j_N}_N\rangle$, where $j_i=1,2$.
The general expression for the 'WWZB-inequalities' is given by a
linear combination of the correlation expectation values,
\begin{equation}\label{complset}
  I_N(\vec c)=\sum_{j_1,\ldots,j_N} c(j_1,\ldots,j_N)
  E(j_1,\ldots,j_N)\leq 1 ,
\end{equation}
where the conditions for the coefficients $\vec c$ can be found in
\cite{WW,ZB}. Using another standard terminology, $I_N(\vec c)$ is
the expectation value of the {\em Bell operator} \cite{BMR}
defined as \be B_N= \sum_{j_1,\ldots,j_N} c(j_1,\ldots,j_N)
O^{j_1}_1\otimes \ldots\otimes O^{j_N}_N\,; \ee the Bell
inequality reads $\mbox{Tr}(\rho B_N)\leq 1$.

The WWZB set of inequalities (\ref{complset}) is complete in the
following sense. If any of these inequalities is violated, the
observed correlations $E(j_1,\ldots,j_N)$ do not admit a LV
description. If none of these inequalities is violated, there
exists a LV model for the list of data $E(j_1,\ldots,j_N)$ --- but
this LV model may not reproduce the correlations of another subset
of observables \cite{WW}.

An important member of this family of inequalities is the
Mermin-Belinskii-Klyshko (MBK) inequality \cite{Mermin,GBP}. Given
that each local observer $i=1,...,N$ measures either $O_i^{1}=\hat
n_i\cdot\vec{\sigma}\equiv \sigma(\hat n_i)$ or $O_i^{2}=
\sigma(\hat n'_i)$, the $N$-qubit Bell operator for these
inequalities is defined recursively as
\begin{eqnarray}\label{mermop}
    M_N&=&\frac{1}{2}\big[(\sigma(\hat n_N)+\sigma(\hat n'_N))
    \otimes M_{N-1} \nonumber\\
    &+&  (\sigma(\hat n_N)-\sigma(\hat n'_N))\otimes
    M'_{N-1}\big] ,
\end{eqnarray}
where $M'_n$ is obtained from $M_n$ interchanging $\hat n_i$ and
$\hat n'_i$, and $M_1=\sigma(\hat n_1)$.

Each inequality in the WWZB family is maximally violated by the
$N$-qubit GHZ state \cite{WW,sca1}. The maximal quantum violation
of the set of inequalities is obtained for the MBK one \cite{WW},
with the choice of measurements optimized for the GHZ state, and
it is equal to $2^{(N-1)/2}$. Then, quantum mechanical violations
of WWZB inequalities range from one up to this value, but not all
the inequalities can be violated up to this bound.

It has also been noticed \cite{GBP,WW3} that in the interval
$]1,2^{(N-1)/2}]$ and for the MBK case one can define {\em degrees
of violation} that are associated to the degrees of entanglement.
Specifically: an $N$-qubit state in which at most $M\leq N$ qubits
are entangled cannot violate the MBK inequality by more than
$2^{(M-1)/2}$. In other words, if the violation exceeds
$2^{(M-1)/2}$, one can be sure that at least $M+1$ qubits are
entangled. In particular, all the $N$ qubits must be entangled in
order to observe a violation in the highest range
$]2^{(N-2)/2},2^{(N-1)/2}]$. Our main result consists on linking
the amount of violation of these inequalities with the degree of
distillability, which is of course a much stronger result than the
link with the degree of entanglement, for all the WWZB
inequalities.

\section{Amount of violation and degree of distillability}
\label{secres}

\subsection{The theorems}

Here we give the results of our investigation. We give them in the
form of two theorems and one corollary. The mathematical proofs
are just sketched and are discussed elsewhere \cite{fulllength} in
all detail.

\begin{center}
\begin{figure}
\epsfxsize=8cm \epsfbox{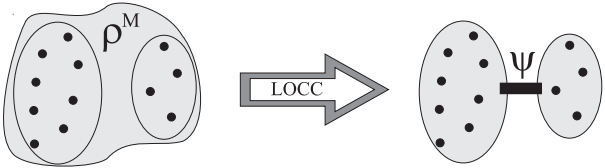} \caption{Theorem 1
illustrated for $N=12$ qubits. The state $\ket{\Psi}$ is a
two-qubit maximally entangled state.} \label{figthm1}
\end{figure}
\end{center}

{\bf Theorem 1.} {\em Any $N$-qubit state $\rho$ that violates one
of the WWZB inequalities is at least bipartite distillable (Fig.
\ref{figthm1}).}

In clear: if one can find a Bell operator $B_N$ in the WWZB family
such that $\mbox{Tr}(\rho B_N)> 1$, then there exists at least one
partition of the $N$ parties into two groups such that the two
groups can distill a singlet. Loosely speaking, Theorem 1 ensures
that in any state that violates a WWZB Bell's inequality there is
some distillable entanglement, although in order to extract it
some parties may need to join.

The proof given in \cite{fulllength} is constructive, i.e. it
provides the distillation protocol. It is a combination of three
known rather technical results: (i) If a state $\rho$ violates a
WWZB inequality, then there exists at least a bipartite partition
such that the partial transpose of $\rho$ with respect to this
partition is negative \cite{WW}; (ii) The eigenvectors of any WWZB
Bell operator are GHZ-like states \cite{sca1}; (iii) There exists
a protocol that allows to depolarize any $N$-qubit state $\rho$
onto a state $\rho'$ that is diagonal in a basis of GHZ-like
states \cite{DC}.

\begin{center}
\begin{figure}
\epsfxsize=8cm \epsfbox{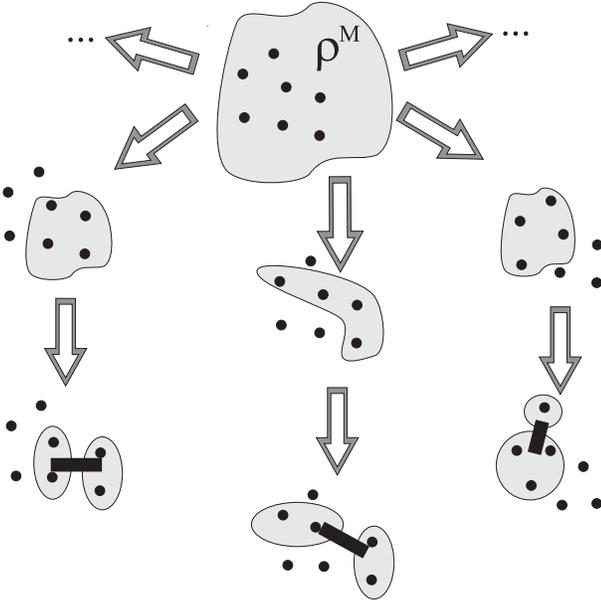} \caption{Theorem 2
illustrated for $N=7$ qubits and a violation measured by $p=4$. As
in Fig. \ref{figthm1}, the hollow arrows represent LOCC
operations, and the thick links represent the singlet state. Any
three $N-p=3$ qubits can perform a suitable measurement and
communicate its result to the others. The four remaining qubits
end up with a state which is bipartite distillable. Note that only
three out of the $C^7_3=7!/(3!\,4!)=35$ possibilities are shown.}
\label{figthm2}
\end{figure}
\end{center}

{\bf Theorem 2.} {\em Suppose that the $N$-qubit state $\rho$
violates one of the WWZB inequalities by an amount of \be
\mbox{Tr}(\rho B_N)\, > \, 2^{\frac{N-p}{2}} \label{condnp} \ee
for a given integer $p$ such that $2\leq p\leq N$. Then {\em any}
ensemble of $p$ qubits can be divided into two subgroups, and a
singlet can be distilled between these subgroups by means of
operations which are local with respect to the $N-p+2$ parties
(Fig. \ref{figthm2}).}

The proof is also constructive. If $p=N$, we have Theorem 1: there
is just one ensemble of $p=N$ qubits, which is the full set, and
we know that if the inequality is violated then the state is
bipartite distillable. Suppose now that $p\leq N-1$. Then one can
rather easily show the following: if any of the observers measures
his qubit in a suitable way and communicates the result to the
others, then the resulting conditional state $\rho_{N-1}$ of $N-1$
qubits is such that (see \cite{fulllength} for details) \be
\mbox{Tr}(\rho_{N-1} \tilde{B}_{N-1})\,\geq\,
\frac{1}{\sqrt{2}}\mbox{Tr}(\rho B_N)\,. \ee Here,
$\tilde{B}_{N-1}$ is a WWZB Bell operator on $N-1$ qubits, linked
in a constructive way to the original $B_N$. By recursive
application of this argument: we start from (\ref{condnp}); if any
$N-p$ parties perform the suitable local measurement and
communicate the result to the others, then the remaining $p$
parties share a conditional state $\rho_p$ that still violates a
WWZB inequality. To these $p$ qubits we apply Theorem 1.

Note that in its statement Theorem 1 is a corollary of Theorem 2;
however, the proof of Theorem 2 that we have just described relies
on the validity of Theorem 1, which is therefore logically
independent in our construction. Let us conclude with an important
corollary of Theorem 2:

{\bf Corollary.} {\em If $N$-qubit state $\rho$ violates one of
the WWZB inequalities in the range \be 2^{\frac{N-2}{2}}\,<\,
\mbox{Tr}(\rho B_N)\, \leq \, 2^{\frac{N-1}{2}} \label{condfull}
\ee then $\rho$ is fully distillable.}

This Corollary can be derived independently by showing that for
(\ref{condfull}) to hold, $\rho$ must have (up to local unitary
transformations) a large overlap with the $N$-qubit GHZ state. In
fact, the bounds given in \cite{fulllength} imply in particular
$\sandwich{\mbox{GHZ}}{\rho}{\mbox{GHZ}}>2/3$ for all $N>3$. When
one refers to the work of D\"{u}r and Cirac \cite{DC}, it is easy
to see that this condition implies full distillability (in the
notation of these authors, $\sandwich{\mbox{GHZ}}{\rho}{
\mbox{GHZ}}=\lambda_0^{+}$).

\subsection{Two comments}

Let us add two comments on the above results.

First, we stress that the bounds given in these Theorems cannot be
improved on the whole WWZB class. Suppose in fact that we just
know that for a state $\rho$ there exists a WWZB Bell operator for
which $\mbox{Tr}(\rho B_N)= 2^{(N-2)/2}$. This state might be
$\ket{\mbox{GHZ}}_{N-1}\otimes\ket{0}$, which is known to reach
this bound for the MBK inequality, and which is clearly not fully
distillable. One might say that the criterion of Theorem 2 is very
simple, since it involves only one number (the amount of
violation), and its roughness is the price to pay for its
simplicity. Certainly, more sharp criteria could be derived if one
has some knowledge of the inequality; and we have shown elsewhere
\cite{fulllength} that if one restricts to some specific
inequalities, then a refinement is indeed possible \cite{note1}.

The second comment concerns bound entanglement. D\"{u}r \cite{Dur}
was the first to notice the existence of states that violate a
WWZB (actually, a MBK) inequality, while being bound entangled for
some partitions (in his case the most ``natural" partition into
$N$ qubits). This result fits well in our general scheme, on which
it casts some new light. In fact, in order to have a violation, we
have shown that it is necessary that {\em some} partitions are
associated to distillable entanglement (see also \cite{acin}),
while D\"{u}r's example shows explicitly that it is not
necessarily the case for all partitions. We would like to stress
in this context a matter of terminology: in multi-partite systems,
the notion of ``bound entangled states that violate an inequality"
is equivocal, precisely because the entanglement may be bound for
some partitions and distillable for other ones. Based on the
results and the ideas described here, we strongly believe that
{\em distillable} entanglement is responsible for the violation of
a Bell's inequality in $N$-qubit systems.

\section{Conclusion and open questions}

In this work, we have discussed the complete solution of the link
between distillability and the WWZB family of Bell's inequalities
for $N$ qubits. This study illustrates the complexity and the
richness of the structure of multi-partite entanglement.

Of course, this work is susceptible of some extension even as far
as the case of $N$ qubits is concerned, since after all we have
proved the link with distillability ``only" for the WWZB
inequalities. A first direction would be to study similar
connections for inequalities with more settings per site
\cite{dag}. Nevertheless, note that one cannot exclude the
possibility that from a quantum state violating a Bell's
inequality that does not belong to this family (with more setting
or outcomes) a new state can always be obtained by LOCC violating
an inequality in the WWZB family. Very little is known about the
structure of Bell's inequality quantum violations under LOCC
transformations.

A hotter related open problem is the extension to higher
dimensional systems. Even in the case of bipartite systems
$\compl^{d}\otimes \compl^{d}$, where there is bipartite bound
entanglement, nothing is known about the link between violation of
Bell's inequalities and distillability \cite{red}. As we said
above, it would come out as a great surprise if non-distillable
states were found to violate some Bell's inequality --- and this
is in itself a good reason to investigate this problem.

We want to conclude on Bell's inequalities. These inequalities,
initially derived to quantify the counter-intuitive features of
quantum correlations, are acquiring the status of ``witnesses for
useful entanglement". It had already been shown that the violation
of a multi-qubit inequality is a sufficient condition for security
of key-distribution protocols \cite{sca2}. A link between Bell's
inequality violation and communication complexity has been also
noticed \cite{commcompl}. Now we have proved that if a $N$-qubit
state violates any of the inequalities found in \cite{WW,ZB}, then
its entanglement can be distilled.

\section{Acknowledgements}

We thank Nicolas Gisin and Wolfgang D\"ur for discussions. We
acknowledge financial support from the Swiss NCCR "Quantum
Photonics" and OFES, within the project EQUIP (IST-1999-11053),
and the DFG. V. S. acknowledges support from an ESF Short
Scientific Visit Grant.

\end{multicols}


\begin{thebibliography}{10}
\bibitem{maxent}
A pure state $\ket{\Psi}\in\compl^d\otimes\compl^d$ is a maximally
entangled state when it can be written as
$\ket{\Psi}=1/\sqrt{d}\sum_{i=1}^d\ket{ii}$ where $\{\ket{i}\}$
are orthonormal local bases. For distillability it is suffcient to
demand that a two-qubit maximally entangled state, also called
singlet, can be obtained by LOCC.
\bibitem{PT}
Consider an operator, $O$, that acts on
$\compl^{d_1}\otimes\compl^{d_2}$. The partial trasposition of $O$
with respect to the first subsystem in the basis
$\{\ket{1},\ldots,\ket{d_1}\}$ is given by $O^{T_1}\equiv
\sum_{i,j=1}^{d_1}\bra{i}O\ket{j}\ket{j}\bra{i}$.
\bibitem{note0} To be precise, bound entangled states do not exist
also for $\compl^{2}\otimes \compl^{3}$.
\bibitem{note1}
Indeed we have examples where an infinitesimal Bell's inequality
violation is sufficient for fully distillability.
\bibitem{red} We note that there are examples of states violating
the CHSH inequality although satisfying the reduction criterion.
The violation of the latter is one of the most powerful calculable
sufficient criteria for bipartite distillablility.
\end{thebibliography}
\end{document}